\documentstyle[11pt,newpasp,twoside,epsf]{article}
\def\edcomment#1{\iffalse\marginpar{\raggedright\sl#1\/}\else\relax\fi}
\reversemarginpar

\begin{document}

\title{Multi-transition VLBA observations of circumstellar SiO masers}
\author{Soria-Ruiz R., Colomer F., Alcolea J., Desmurs J.-F., Bujarrabal V.}
\affil{Observatorio Astron\'omico Nacional, Spain}
\author{Marvel, K.M.}
\affil{American Astronomical Society}
\author{Diamond, P.J.}
\affil{Jodrell Bank Observatory}
\author{Boboltz, D.}
\affil{US Naval Observatory}

\begin{abstract}
On this paper, we present simultaneous VLBA observations of the $J$=1--0 
and $J$=2--1 rotational lines in the $v$=1 and $v$=2 vibrationally excited 
states of SiO (at $\lambda$\,$\sim$\,7 and 3~mm). We have mapped these four
maser lines in the circumstellar envelopes  of three AGB stars: IRC\,+10011, 
$\chi$\,Cyg and TX\,Cam. We study the relative spatial 
distribution between these maser lines. In particular, for the two $v$=1 
transitions, we found that the $J$=1--0 and $J$=2--1 maser spot distributions
are unalike, challenging all the current theoretical pumping mechanisms for
SiO masers (which predict quite similar distribution for maser lines in the
same vibrational excited state).
\end{abstract}

\section{Introduction}
General conclusions can be derived from the current theoretical pumping mechanisms of the SiO masers, as the production of masers chains across the same vibrational state, the time variability of the maser features or the location of the SiO layer in the circumstellar envelope. The main difference between the radiative (e.g. Bujarrabal 1994a; 1994b) and collisional (e.g. Locket \& Elitzur 1992; Humphreys et al. 2000; 2002) models rests on which is the primary mechanism responsable for the pumping of these masers, which is the IR stellar radiation in the first case and the collisions with the circumstellar gas particles in the second.
  
We have conducted SiO masers observations in order to compare the different predictions of the theoretical pumping models with the observational results. In particular, we focussed on the comparison of the spatial distribution between the $v$=1 $J$=1--0 and $v$=2 $J$=1--0, and the $v$=1 $J$=1--0 and $J$=2--1 rotational transitions. 

\section{Observations}

We have performed sub-m.a.s. resolution observations of the $J$=1--0 and 2$-$1
rotational maser lines of $^{28}$SiO, at 7mm ($\sim$ 43 GHz) and 3mm (86 GHz) 
respectively, in the $v$=1 and 2 vibrationally excited states. We have
observed  three long period variables: IRC\,+10011, $\chi$\,Cyg and TX\,Cam. 
In table\,1 we summarize some characteristics of these sources and
the detected transitions in each of them. The stellar phase, $\phi$, 
at the time of the observation is given with respect to the optical maximum.
The asterisk marks lines detected that could not be mapped.
At 43~GHz the recorded bandwidth was 8~MHz with a spectral resolution of 
0.22~km\,s$^{-1}$, while at 86~GHz we use a 16~MHz bandwidth with a
resolution of 0.11~km\,s$^{-1}$.
The integrated intensity maps were obtained following the standard calibration procedure for spectral lines within the AIPS package. 
\vspace{-0.2cm} 
\begin{table}
\caption{Observed sources and SiO transitions}
\vspace{0.2cm}
\begin{tabular}{|cccc|cc|cc|}
\hline
Source & Variability & Mass loss & $\phi$ & \multicolumn{2}{c|}{43 \small{GHz}}&\multicolumn{2}{c|}{86\small{ GHz}}\\
& type&(M$_{\odot}$\,yr$^{-1}$) &&$v$=1&$v$=2 & $v$=1& $v$=2\\
 \hline
 \hline
IRC +10011 & OH/IR & \small{$\sim$}\,8.5\,$10^{-6}$ &\small{ 0.10} & det. &det.&det.&non-det.\\
$\chi$ Cyg & Mira & \small{$\sim$}\,5.6\,$10^{-7}$ &\small{ 0.25} &det.& det.$^{*}$ &det. &det.$^{*}$ \\
TX Cam & Mira& \small{$\sim$}\,2.5\,$10^{-6}$ & \small{0.50} & det.& det. &det.$^{*}$ &non-det.\\ 
 \hline
\end{tabular} 
\end{table} 

\subsection{IRC +10011}

In Fig.\,1, we present the VLBA maps of the two 7\,mm wavelength tansitions, $v$=1 and $v$=2 $J$=1--0 (rest frequencies  of 43.122~GHz and 42.820~GHz respectively).
For these two 7\,mm lines, the spatial distribution of the spots is quite similar, but the sizes of the emitting regions slightly differ. We measure a diameter of $\sim$\,20 m.a.s. for the $v$=1 extent and $\sim$\,18 m.a.s. for the $v$=2. 

\begin{figure}[!hbp]
\plottwo{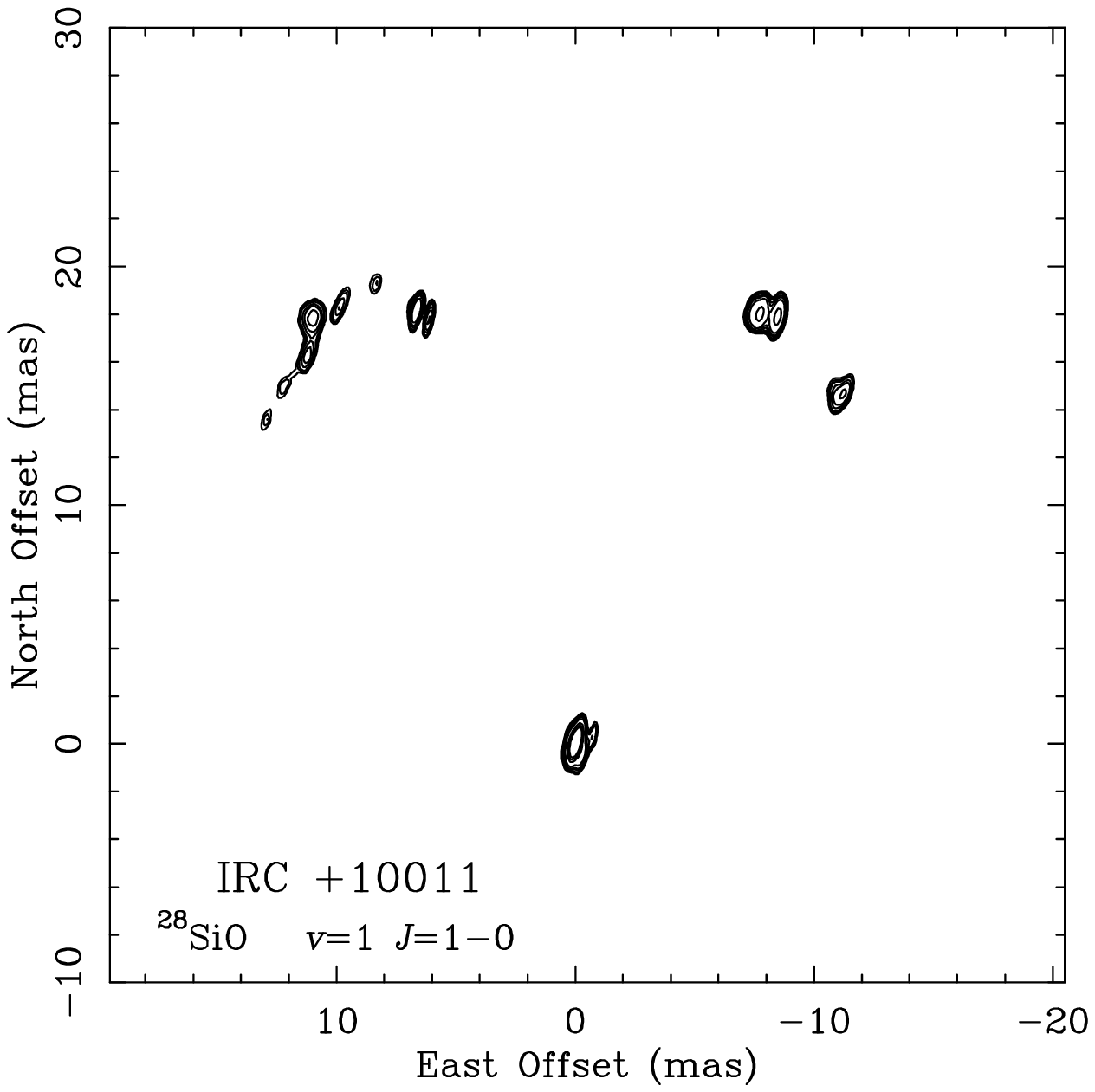}{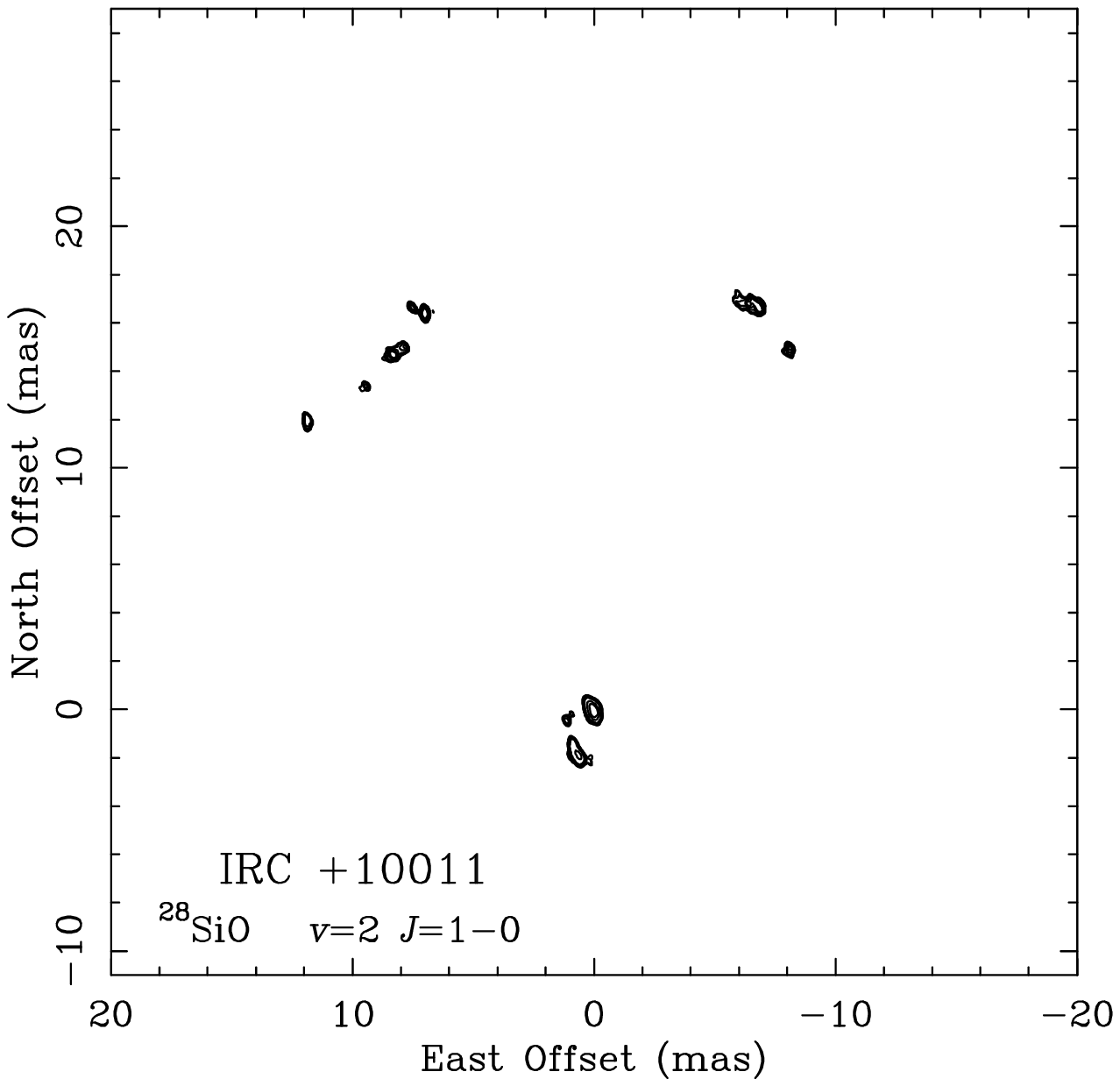}
\caption{Left: VLBA map of the $v$=1 $J$=1--0 maser line for IRC\,+10011. Right: idem for the $v$=2 $J$=1--0 transition.} 
\end{figure}

The 3\,mm VLBA map of the $v$=1 $J$=2--1 maser (86.243 GHz) is shown in Fig.\,2. 
We also present, for the $v$=1 $J$=1--0 and $J$=2--1 lines, the autocorrelation spectrum of the reference antenna, the cross-correlation or flux recovered after the calibration and imaging, and their flux ratio. We note that at 3\,mm, almost the 90\% of the flux, as an average, has been lost, while at 7\,mm the recovered flux is about a 40\% at velocities near the systemic velocity (dotted lines).   

The $v$=1 $J$=2--1 transition presents a completely different distribution compared to the $v$=1 $J$=1--0, the emitting region being also significantly larger, of about 25--30 m.a.s. (Note that all maps are shown to the same scale to ease a direct comparison).

\begin{figure}[!htp]
\plottwo{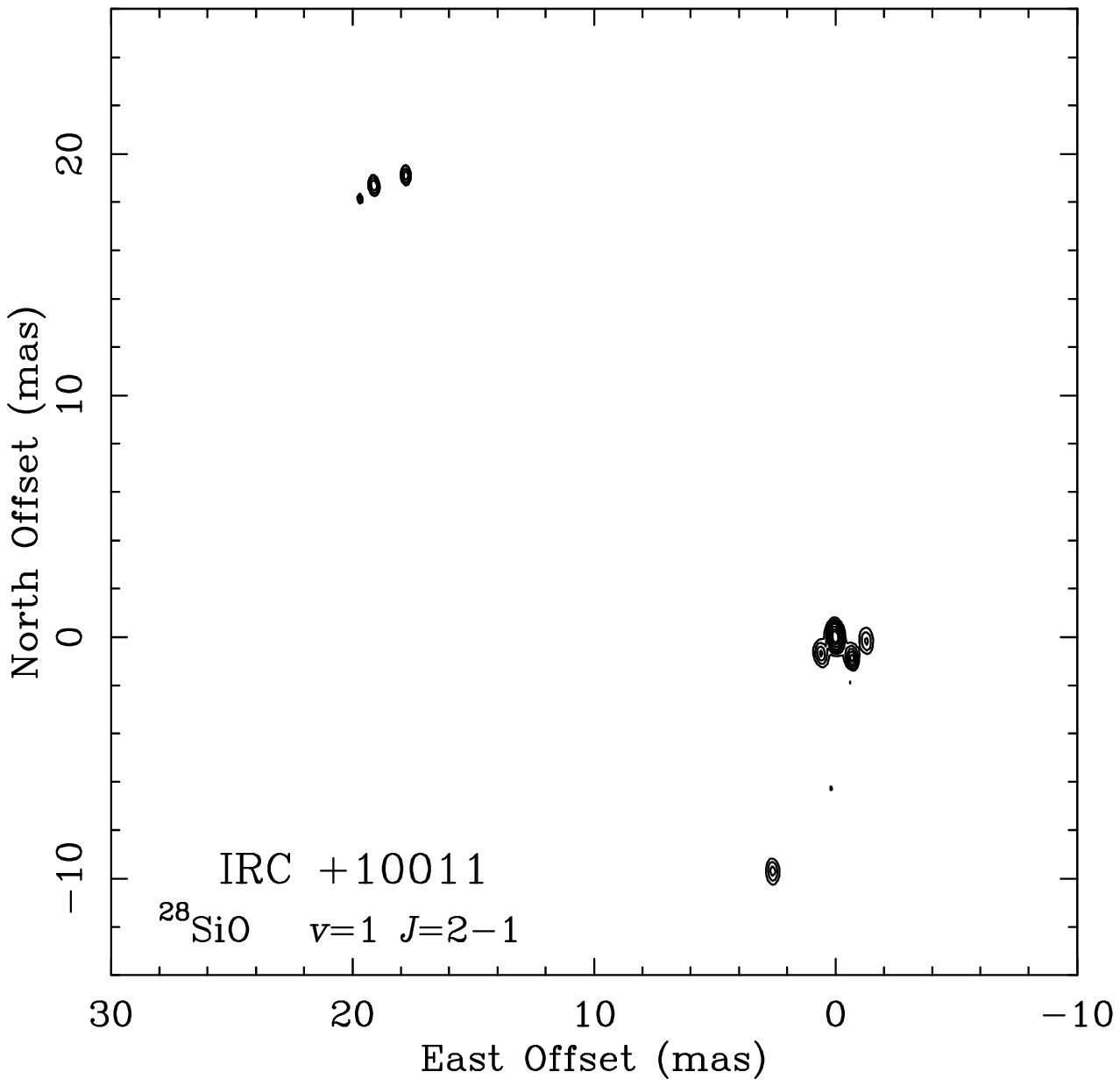}{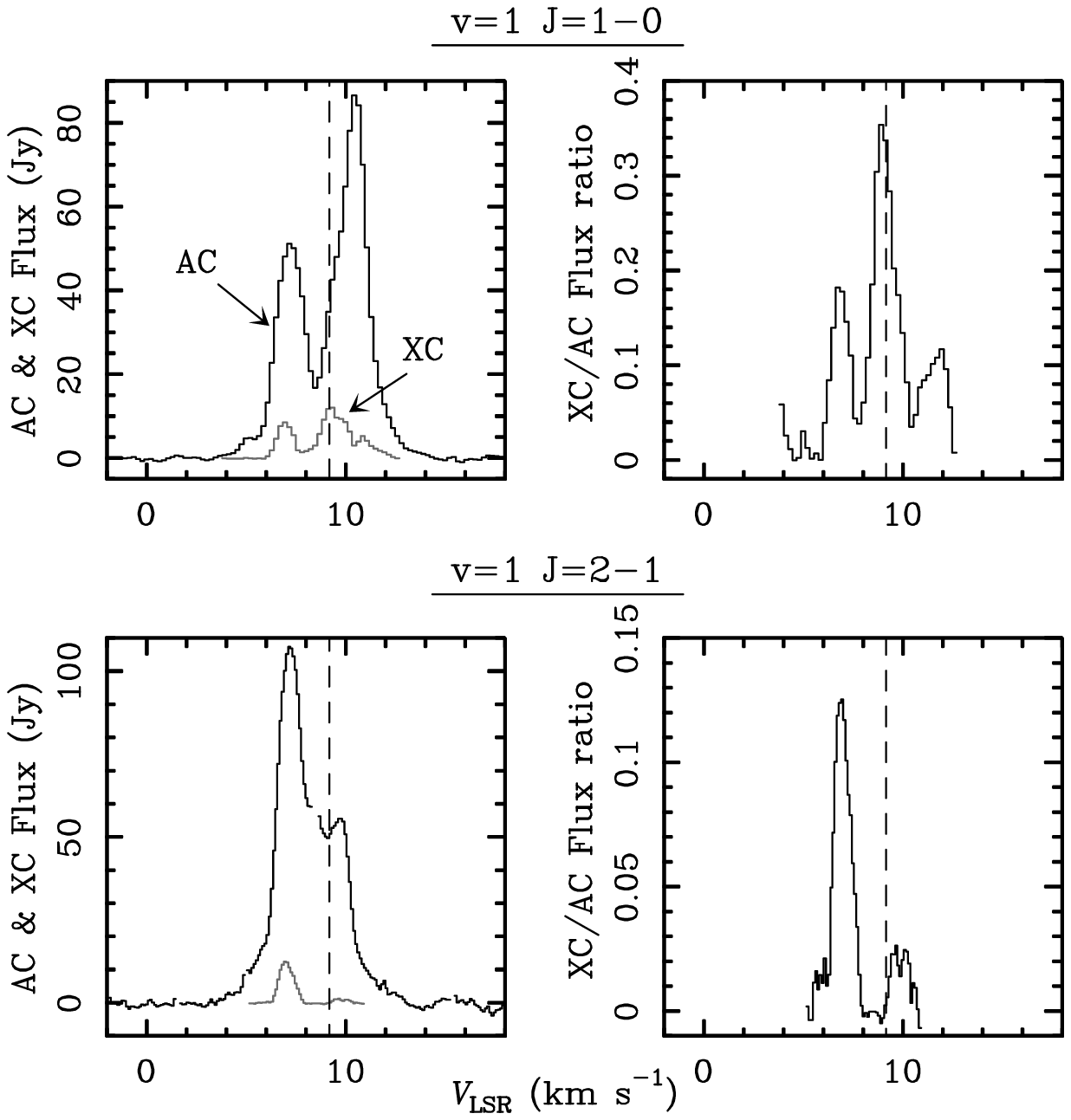}
\caption{Left: VLBA map of the $v$=1 $J$=2--1 (86.243 \small{GHz}) transition for IRC\,+10011. Right: Autocorrelation and crosscorrelation spectra, and XC/AC-ratio for the $v$=1 $J$=1--0 and $v$=1 $J$=2--1 transitions.} 
\end{figure}

\subsection{$\chi$ Cyg}

The four $^{28}$SiO transitions observed (table 1) were detected in $\chi$ Cyg although only the $v$=1 $J$=1--0 and the $v$=1 $J$=2--1 are presented in figure 3.
The $v$=2 $J$=1--0 spectrum consists in a single feature, and consistently the map presentes an unresolved single spot.
For the  $v$=2 $J$=2--1 transition, imaging was not possible because of the low data quality.       

\begin{figure}[!h]
\plottwo{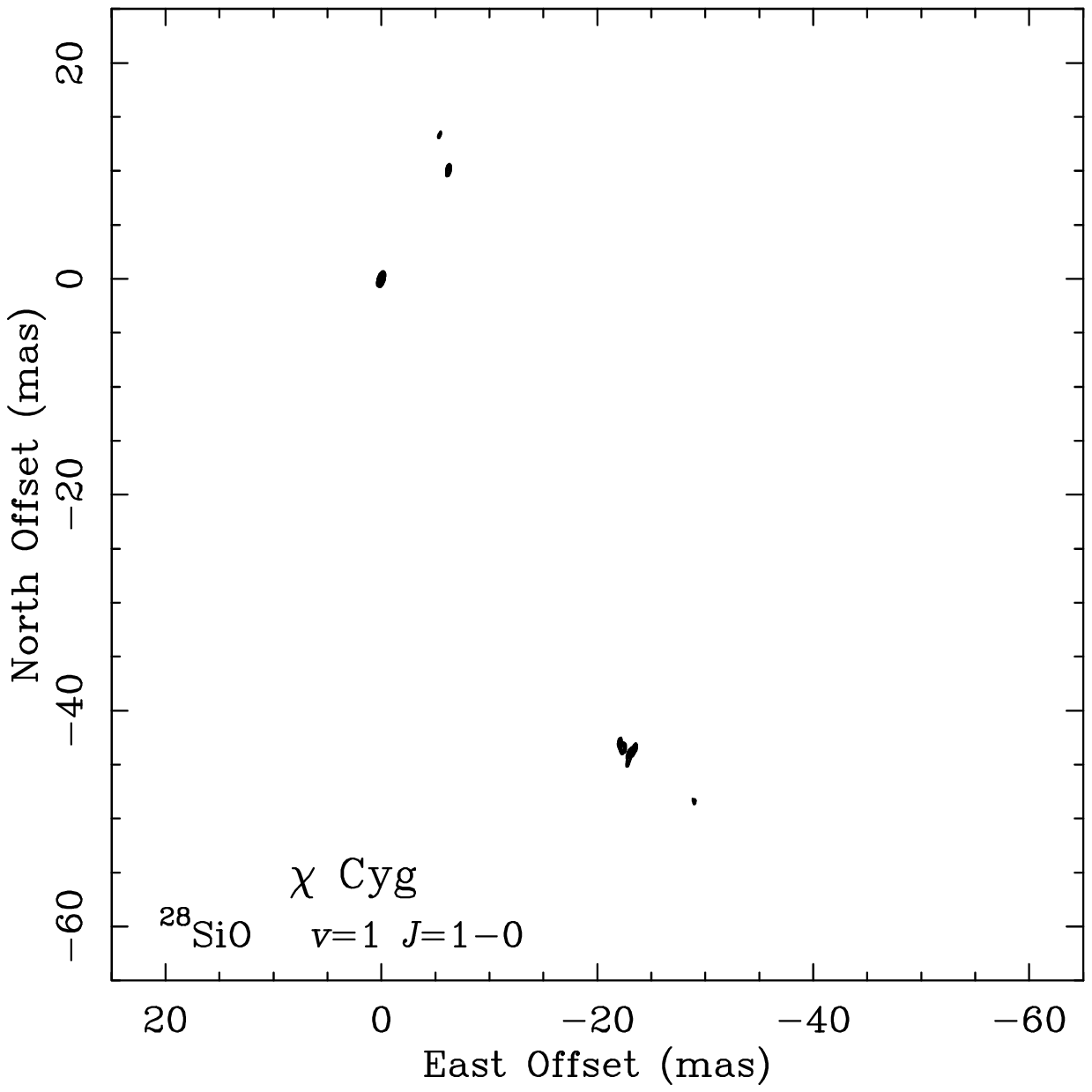}{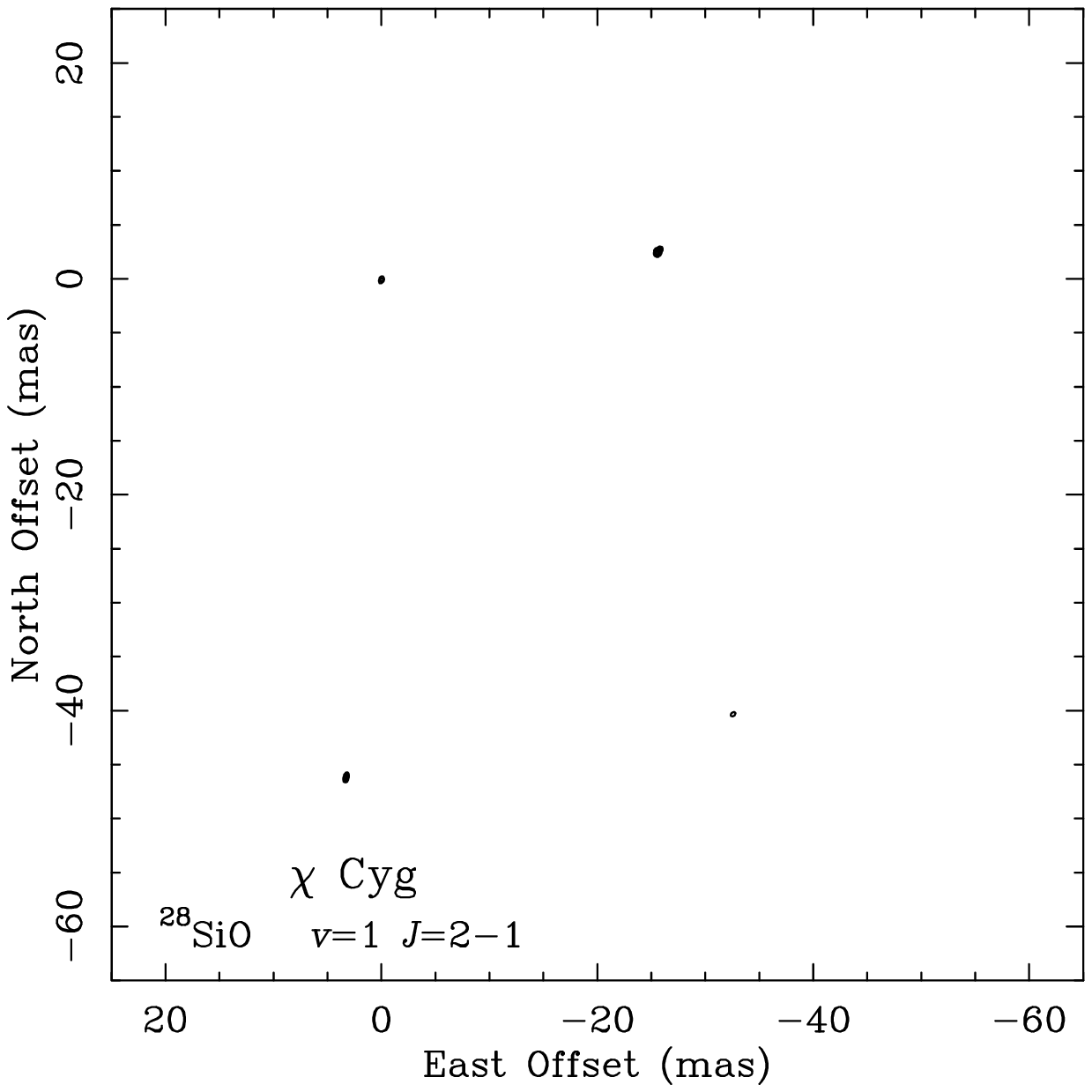}
\caption{Left: VLBA map of the $v$=1 $J$=1--0 maser line for $\chi$ Cyg. Right: idem for the $v$=1 $J$=2--1 transition.} 
\end{figure}

The comparison of the $v$=1 VLBA maps of $\chi$ Cyg again shows a totally different distribution for these lines. However, in this case, the emitting regions have comparable sizes, $\sim$\,45--47 m.a.s. for the $J$=1--0 and $\sim$\,45 m.a.s. for the $J$=2--1, in contrast with the results for IRC\,+10011.

\subsection{TX Cam}

We found no emission at the 3\,mm transition $v$=2 $J$=2--1, and for the $v$=1 $J$=2--~1 we could not calculate the delays as the calibrator was not detected, and so only the 7\,mm comparison was possible. The maps of the two $J$=1--0 SiO masers are presented in figure 4.

\begin{figure}[!h]
\plottwo{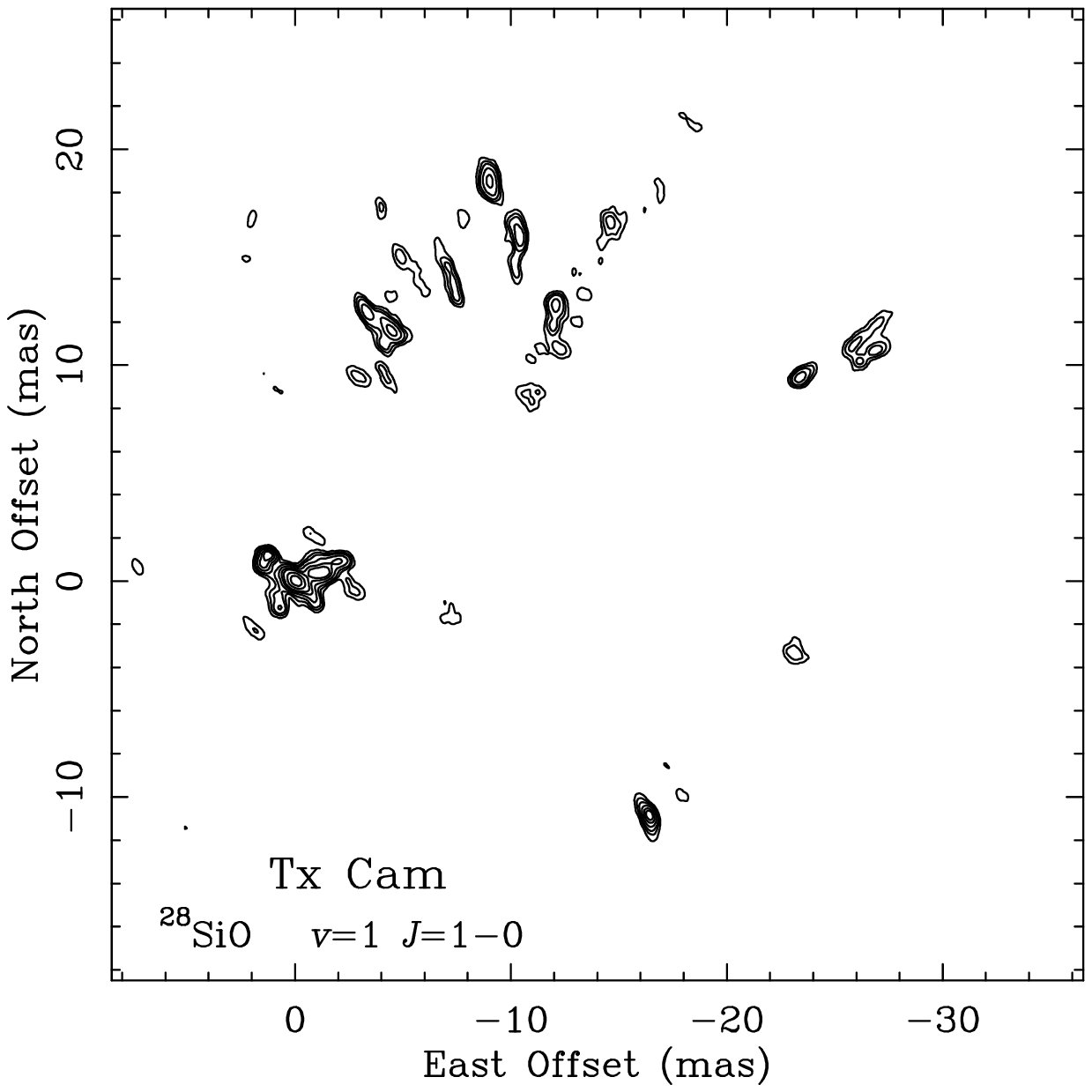}{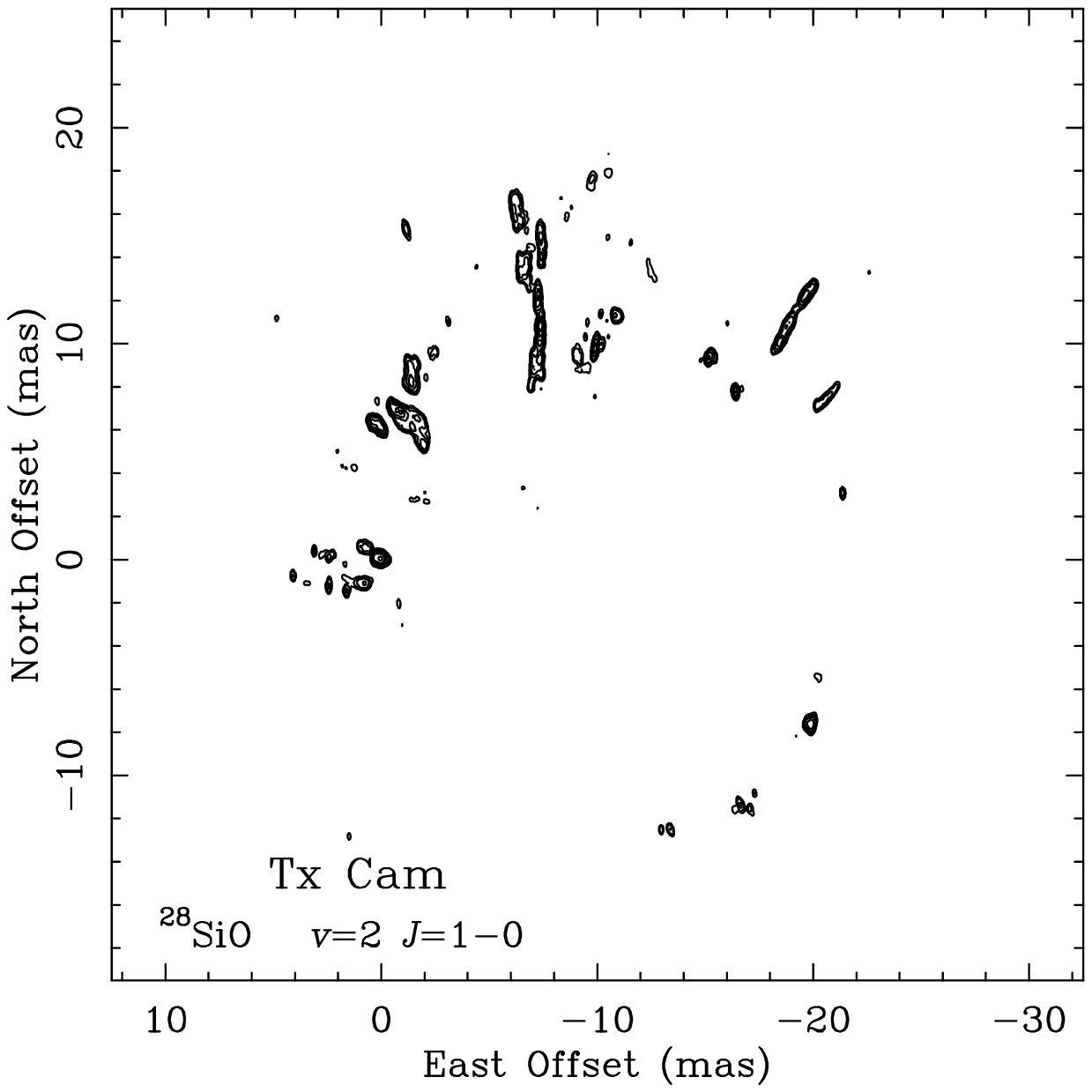}
\caption{Left: VLBA map of the $v$=1 $J$=1--0 maser line for TX~Cam. Right: idem for the $v$=2 $J$=1--0 transition.} 
\end{figure}

The spatial distribution of the spots for both transitions is similar. The $v$=2 $J$=1--0 emission appears to be located over a ring structure in a slighter inner region than the $v$=1 $J$=1--0 line. The diameter sizes are $\sim$\,26 m.a.s. and $\sim$\,28 m.a.s. respectively, taking into account only the brightest spots. (Note the radial enlongated shape of some features in both maps).   

\section{Conclusions and work in progress}

Theoretical models predict that in the case of the 7\,mm transitions, i.e., the $v$=1 $J$=1--0 and $v$=2 $J$=1--0, the ring-like structure sizes are similar (Herpin 1998) or slightly different, appearing to be the $v$=2 $J$=1--0 emission distributed in an inner region (Humphreys et al. 2000). In the case of the 7\,mm and 3\,mm comparison, i.e., the $v$=1 $J$=1--0 and $J$=2--1, all models predict that both transitions have similar spatial distributions and ring sizes (Locket \& Elitzur 1992; Bujarrabal 1994a; Humphreys et al. 2002), as a natural consequence of their mutual reinforcement. 

For the $v$=1 and $v$=2 $J$=1--0 comparison we have found for IRC\,+10011 (figure 1) and TX Cam (figure 4) a similar spatial distribution of the spots of both transitions, being the $v$=2 emitting region in an inner region, in agreement with Humphreys et al. (2000). We confirmed the results already measured by Desmurs et al. (2000) in the same two sources. (Note that for $\chi$ Cyg, altough the line was detected, only one spot appeared in the $v$=2 $J$=1--0 map so we could not compare its spatial distribution to the $v$=1 one).

However, for the $v$=1 $J$=1--0 and $J$=2--1 map comparison, we have found in our results a contradiction with the models, since, in the case of IRC\,+10011, we have a different distribution and emitting regions sizes, the $J$=2--1 maserspots lying outside the $J$=1--0 ones (figure 1 left and figure 2). For $\chi$\,Cyg, we have obtained similar sizes (figure 3) but a totally different spatial distribution. (A similar study has been recently published by Phillips et al. (2003) where however they found a similar distribution and ring-like structure sizes for the 7\,mm and 3\,mm maser lines, towards the M-type star R Cas).      

Our results are not very surprising since single-dish observations demonstrate that the profiles of the $J$=1--0 and $J$=2--1 transitions are often very different (Lane 1982), so unlike spatial distributions are expected.
The contradiction with the models' predictions could be due to effects not taken into account in the calculations like line overlapping, which is known to play an important role in the pumping of the $v$=2 $J$=2--1 line and most of the $^{29}$SiO and $^{30}$SiO (see e.g. Bujarrabal et al. 1996 and Gonz\'alez-Alfonso 1997).   
Presently, we are investigating the prevalence of our results on a second epoch already observed with the VLBA, in the same sources and in R Leo. Further theoretical development on pumping models is in progress.

\end{document}